\documentclass[
    ,final            % use final for the camera ready runs
  ]
  {aipproc}

\layoutstyle{6x9}

\begin{document}

\title{Light-like Wilson Loops and Cusp Anomalous Dimensions in Nonconformal Gauge Theories}

\classification{11.25.-w,11.25.Tq,11.15.Me}
\keywords      {Gauge/string duality, Strong-coupling expansions, Wilson loops}

\author{Leopoldo A. Pando Zayas}{
  address={{Michigan Center for Theoretical
Physics}\\
{ \it Randall Laboratory of Physics, The University of
Michigan}\\
{\it Ann Arbor, MI 48109-1040}}}

\author{Daniel J. Phalen}{
  address={{Michigan Center for Theoretical
Physics}\\
{ \it Randall Laboratory of Physics, The University of
Michigan}\\
{\it Ann Arbor, MI 48109-1040}}
}

\author{C\'esar A. Terrero-Escalante}{
  address={{Centro de Investigación y Estudios Avanzados}}\\
{\it Av. Polit\'ecnico Nacional 2508, M\'exico D.F. 07360, M\'exico}}

\begin{abstract}
We emphasize that nonconformal theories provide a natural playground for the ideas of the 
Maldacena conjecture opening the possibility of exploring properties that could potentially be in the 
same universality class as  QCD. In particular, we discuss in detail how light-like Wilson loops, 
an important ingredient in the prescription for scattering amplitudes, can be described in a number of gravity 
duals of nonconformal gauge theories.  We point out to a few universal properties and the prominent role of the 
strong scale.
\end{abstract}

\maketitle

%%%%%%%%%%%%%%%%%%%%%%%%%%%%%%%%%%%%%%%%%%%%
%% MAINMATTER
%%%%%%%%%%%%%%%%%%%%%%%%%%%%%%%%%%%%%%%%%%%%

\section{Introduction}

The idea that the description of gauge theories 
might effectively involve strings has a long history starting from the dual models for hadronic 
resonances. In 1974 `t Hooft \cite{'t Hooft:1973jz} showed that gauge theories admit a limit in which  their perturbation 
theory can be thought of as coming from a string theory. 
Ten years ago, a precise formulation of this equivalence was put forward for a particular gauge theory. Namely, it 
was argued that ${\cal N}=4$ supersymmetric Yang-Mills (SYM) is 
dual to type IIB string theory propagating on $AdS_5\times S^5$ \cite{Maldacena:1997re,Witten:1998qj,Gubser:1998bc}. 
In the ten years 
following the seminal work of Maldacena \cite{Maldacena:1997re}
the gauge/gravity correspondence has seen a period of maturity.  
In its current incarnation the correspondence has turned into a very fruitful theoretical framework. 

Particularly exciting is the side of Maldacena's conjecture relating the strongly 
coupled sector of the gauge theory with weakly coupled gravity.
Many problems in gravity have been solved during 
nearly a century of work. By turning strong coupling problems of gauge theories into questions in classical gravity, the 
correspondence opened a window of opportunity into one of the 
long-standing puzzles of particle physics: the infrared properties of gauge theories.
Some of the questions to be addressed here are confinement, the spectrum 
of hadrons and the properties of the strongly coupled quark-gluon plasma. However, important questions 
about nonprotected quantities lie beyond the domain of classical gravity and should properly be 
tackled using string theory. 

There are two main obstacles 
in applying the gauge/gravity correspondence to phenomenologically interesting theories. 
The high amount of supersymmetry and conformal symmetry is not present in nature. 
Although originally formulated as a duality between string 
theory on $AdS_5\times S^5$ and maximally SYM in four 
dimensions, it has quickly become clear that such dualities extend to more general situations. 
Significant results in understanding QCD-like theories 
have been achieved by constructing supergravity theories whose dual gauge theories 
contain ${\cal N}=1$, $SU(N)$ supersymmetric Yang-Mills \cite{Maldacena:2000yy,Klebanov:2000hb}. 
In a remarkable series of papers \cite{kw} Klebanov and collaborators carried
out a program that concluded with a supergravity background that is dual to ${\cal N}=1$ supersymmetric 
gauge theory that displays confinement and chiral symmetry breaking in the IR. 
This background is known as the
warped deformed conifold or the Klebanov-Strassler (KS) solution.
In the UV this background encodes a series of Seiberg duality cascades in the field theory. 
A fairly comprehensive description of how to encode ${\cal N}=1$ gauge dynamics into supergravity 
backgrounds was presented by Sonnenschein and Loewy in \cite{Loewy:2001pq}.
They analyzed the following properties: 
gauge group, supersymmetries, 
Wilson and `t Hooft loops and the corresponding quark anti-quark or monopole anti-monopole potentials, 
instantons and the $U(1)_R\rightarrow Z_{2N}$ symmetry breaking, 
gluino condensation and spontaneous $Z_{2N} \rightarrow Z_{2}$ breaking, monopoles, 
domain walls, 
baryons and
KK states.
Moreover, a supergravity solution dual to the finite temperature warped conifold with cascading in the UV was found \cite{PandoZayas:2006sa}. 
Evidence of a transition between this solution and the KS solution was reported in \cite{Mahato:2007zm} 
(see also \cite{Aharony:2007vg}). 
Such transition could be interpreted as the gravity dual to the confinement/deconfinement transition in QCD-like theories. 

Another important direction in which the Maldacena conjecture has been expanded is in going beyond the lowest energy states. 
Much of the developments in the conformal version regarding 
beyond the supergravity approximation were extended to the nonconformal case.
For example, the BMN limit \cite{Berenstein:2002jq} was also understood in nonconformal situations; 
the analogous 
structure describes hadronic states for which an exact string Hamiltonian was presented \cite{Gimon:2002nr}.
Using semiclassical 
quantization in the context of gauge/gravity correspondence, 
quantum corrections to Regge trajectories were calculated \cite{PandoZayas:2003yb}, 
showing that quantum effects alter both the linearity of the trajectory and the vanishing classical intercept, 
$J:=\alpha(t)= \alpha_0 + \alpha' t +\beta\sqrt{t}$.

More recently, Alday and Maldacena proposed a prescription to calculate the gluon scattering 
amplitudes for ${\cal N}=4$ using beyond the supergravity degrees of freedom \cite{Alday:2007hr}. 
Further research in that direction has already gained momentum  
(for instances and references see the contribution by Alday and Maldacena to this volume).
The logical next step is
to apply their prescription to nonconformal situations.
However, several obstacles arise. 
First of all, we have to deal with less symmetry, meaning that we need to use more powerful 
techniques to solve the corresponding problems. 
Fortunately, these problems reduce to systems of ordinary and partial differential equations, and 
we can use many analytical and numerical methods for studying and solving them.

One goal of this contribution is to display the power of these methods.
Using as illustrative examples the Witten-QCD \cite{Witten:1998zw} and KS \cite{Klebanov:2000hb} models, 
we are going to show that analysis of the light-like Wilson loop in the dual momentum space for 
 nonconformal situations  shows some universal properties. Based on the recognizable universality that we obtain,  we 
expect that the analysis of gluon scattering in nonconformal theories would also yield valuable results 
potentially in the same universality class as QCD. 

%$\clubsuit\clubsuit$
%{\it Spin chain and integrability has not found its true place in the context of nonconformal gauge/gravity correspondence. }
%$\clubsuit\clubsuit$

%$\clubsuit \clubsuit$
%{\it The true question remains: From confinement to asymptotic freedom.}
%$\clubsuit \clubsuit$

%%%%%%%%%%%%%%%%%%%%%%%%%%%%%%%%%%%%%%%%%%%%%%%%%%%%%%%%%
\section{Wilson loops in gauge theories}
%%%%%%%%%%%%%%%%%%%%%%%%%%%%%%%%%%%%%%%%%%%%%%%%%%%%%%%%%%%

The Wilson loop is a very important observable in gauge theories and in some approaches it is treated as 
the central ingredient in the description of all gauge invariant quantities.
Its description on the gravity side was proposed in \cite{Maldacena:1998im,Rey:1998ik} where 
it was suggested that the evaluation of the vacuum expectation value of a given Wilson loop on the 
contour ${\cal C}$ is given by the 
string worldsheet whose boundary coincides with ${\cal C}$: $\langle W({\cal C})\rangle = \exp(-S({\cal C}))$. 
\begin{figure}
  \includegraphics[height=.3\textheight]{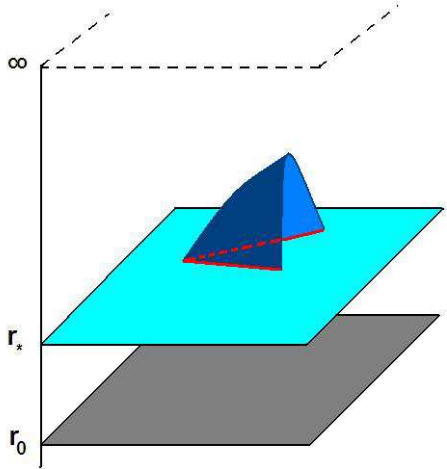}
  \caption{Schematic representation of the Wilson loop.}
\label{fig:Wl}
\end{figure}

The Wilson loop provided the first example of the correspondence that goes beyond the study of protected 
quantities at the supergravity level. 
%In this case one considers a classical string configuration in 
%the given supergravity background. 
Related developments allowed to understand that certain non-BPS operators can be described by 
classical string configurations  in a given supergravity background. For example, the large anomalous 
dimension of operators 
of the form ${\mathop{\rm Tr}} X^I\nabla_{(\mu_1} \ldots \nabla_{\mu_s)}X^I$, where 
$X^I$ are  scalar fields in ${\cal N}=4$ SYM,
%supersymmetric Yang-Mills, 
was computed in \cite{Gubser:2002tv} by considering a large closed string spinning in $AdS_5$, 
\begin{equation}
\Delta -S =\frac{\sqrt{\lambda}}{\pi}\ln S\, ,
\end{equation}
where $\lambda = g_{YM}^2N$ is the 't Hooft coupling. 
These operators are generalizations in ${\cal N}=4$ SYM of the twist-two operators that play a central role in the 
deep inelastic scattering in QCD \cite{Georgi:1951sr,Gross:1974cs}. 
They appear in the operator-product expansion 
of two electromagnetic currents 
$ F_{\alpha\mu_1}\nabla_{\mu_2}\ldots \nabla_{\mu_{s-1}}F^\alpha{}_{\mu_s}$ and 
$\bar{\psi}\gamma_{\mu_1}\nabla_{\mu_2}\ldots \nabla_{\mu_{s}}\psi$. 
Soon after \cite{Gubser:2002tv},  a description of these operators was found in terms of 
light-like Wilson loops in the AdS/CFT correspondence \cite{Kruczenski:2002fb,Makeenko:2002qe}.
One important property of the anomalous dimensions of these operators is that it is conjectured to be 
in general of the form $\Delta-S=f(\lambda)\ln S$, where $f(\lambda)$ is the cusp anomalous dimension 
and it is a function only of $\lambda$.

The Alday-Maldacena prescription for gluon scattering in ${\cal N}=4$ SYM \cite{Alday:2007hr}  
involves computations of 
classical worldsheets in $AdS_5\times S^5$ that end on light-like segments. These configurations are 
light-like Wilson loops.  It is crucial that the space where these Wilson loops live 
is the T-dual space to the original supergravity background.  
This construction motivates us to revisit and expand the analysis of worldsheet configurations ending on 
light-like segments in supergravity backgrounds dual to nonconformal theories. 
Similar configurations have been considered before but with two main differences from those we present here. First, they 
were considered in the original supergravity background, that is, without performing a T-duality and second the 
light-like segments were always placed in the asymptotic UV region. For example, the light-like Wilson loop 
on the Klebanov-Strassler 
background was analyzed in  \cite{Kruczenski:2003wz}. 

In this contribution we report on the situation where we perform a T-duality on the ${\bf R}^{1,3}$ subspace where 
the field theory lives, and the light-like segments are placed in the region 
corresponding to the IR of the field theory. Since we work on the T-dual space and place the brane in an arbitrary position, 
the value of the cusp depends on the structure of the infrared region. 
This is a crucial difference with perturbative QCD where no access to the IR region is possible other than on dimensional grounds. 
%Perturbative QCD techniques are not applicable in the IR region. 

%Here we present some analytical and numerical results for the cusp anomalous dimension in non-conformal gauge theories.
%We focus mostly on WQCD, but also hint the results for
%the Klebanov-Strassler background which is substantially more involved due to its seminumerical nature. 

%%%%%%%%%%%%%%%%%%%%%%%%%%%%%%%%%%%%%%%%%%%%%%%%%%%%%%%%%%%%%%
\subsection{Cusps anomalous dimension}
%%%%%%%%%%%%%%%%%%%%%%%%%%%%%%%%%%%%%%%%%%%%%%%%%%%%%%%%%%%%%%%%%%

If a Wilson loop is evaluated over a closed contour that forms a light-like cusp with angle $\gamma$ in Minkowski space-time, 
then its expectation value is \cite{Polyakov:1980ca}, 
\begin{equation}
W_\gamma \sim \left(\frac{\Lambda}{m}\right)^{-\Gamma_{cusp}(\gamma)}\, ,
\label{eq:Wcusp}
\end{equation}
where $\Lambda$ is a UV cutoff, $m$ is an IR cutoff,
and $\Gamma_{cusp}$ measures the anomalous dimension of the Wilson loop.
The cusp anomalous dimension is also an important observable in gauge theories.
It controls the scaling behavior of various gauge invariant quantities like the logarithmic scaling of the anomalous dimension of higher-spin Wilson operators, 
%(twist-two story was mentioned before but now it is true for any gauge theory),
double-log (Sudakov) asymptotics of elastic form factors in QCD,
the gluon Regge trajectory,
infrared asymptotics of on-shell scattering amplitudes,
%finally, $\Gamma_{cusp}$ 
and it is important for resumming the effects of soft gluon emission in the study of QCD at colliders 
\cite{Korchemsky:1992xv}.

The cusp anomalous dimension depends only on the coupling constant and its expansion at 
weak coupling is known in QCD to three loops \cite{Vogt:2004mw}
 and in ${\cal N}=4$ SYM to four loops  \cite{Bern:2008ap}. In the strong 
coupling regime of ${\cal N}=4$ SYM it was calculated to lowest order 
in \cite{Gubser:2002tv,Kruczenski:2002fb, Makeenko:2002qe}, and to higher orders in \cite{Kruczenski:2007cy,Roiban:2007dq}.
In the strong coupling regime of cascading theories the first results were obtained in \cite{Kruczenski:2003wz}.
For our results we needed to generalize this last approach by putting the brane in a different position. 
In the future we plan to calculate the scattering amplitudes and at that point we 
actually will require a surface instead of an infinitely extending cusp, that is, the surface for scattering ends on 
cusps and thus the surface does not extend to infinity. 

In this note we report mostly on work performed for WQCD and the 
KS backgrounds. 
However, for the strategy of understanding which precise properties of nonconformal theories are universal and which 
are model-dependent, 
it is important to discuss various cases. 
We are currently analyzing also the cases of the nonconfining cascading solution of \cite{PandoZayas:2006sa}, 
the noncritical background ${\bf R}^{1,3}\times SL(2,{\bf R})/U(1)$ \cite{Ashok:2005py} and 
the noncritical solution with an infrared fixed point \cite{Klebanov:2004ya}.
%%%%%%%%%%%%%%%%%%%%%%%%%%%%%%%%%%%%%%%%%%%%%%%
\subsubsection{The general set up for the cusp}

Let us consider a general metric of the form: $ds^2 =q^2(r)dx^\mu dx_\mu + p^2(r) dr^2$.
%\label{eq:backgr}
We perform a T-duality along the $x^\mu$ coordinates and arrive at a metric of the form 
\begin{equation}
ds^2 =\frac{1}{q^2(r)}dy^\mu dy_\mu + p^2(r) dr^2.
\end{equation}
We now consider embedding a string world sheet into this metric. For world sheet coordinates $(\tau, \sigma)$, 
this embedding is 
$y_0=e^\tau \cosh\sigma$, $y_1=e^\tau \sinh \sigma$, $r=r(\tau,\sigma)$ .
%\begin{equation}
%y_0=e^\tau \cosh\sigma, \qquad y_1=e^\tau \sinh \sigma, \qquad %r=r(\tau,\sigma)\, .
%\end{equation}
The action is thus
\begin{eqnarray}
S&=&\frac{1}{2\pi \alpha'} \int d\sigma d\tau \frac{e^{2\tau}}{q^2}\sqrt{1-p^2q^2 e^{-2\tau} \left((\partial_\tau r)^2 
-(\partial_\sigma r)^2\right)}.
\label{eq:Sg}
\end{eqnarray}

The methodology is then to derive the equation of motion for $r$, to solve it, 
substitute the solution into (\ref{eq:Sg}) and integrate for the action.
This integral typically diverges, so an UV cutoff must be chosen as upper limit of integration. Given that $S=\ln W$, 
%\begin{equation}
%S=\ln W \, ,
%\end{equation}
the cusp anomalous dimension can be obtained from (\ref{eq:Wcusp}) as,
\begin{equation}
\Gamma_{cusp}=-\frac{1}
{W}\frac{\partial W}{\partial \ln \Lambda}\, 
= 
-\frac{\partial S}{\partial \ln \Lambda}\,.
\label{eq:cusp}
\end{equation}

\subsubsection{The cusp in ${\cal N}=4$ SYM}

Here we present elements of a systematic analysis to the derivation of the solution for this most symmetric case. 
It might seem like an overkill, however,
reproducing analytical results for simple cases helps to adjust the numerical methods for handling
less symmetric situations. Here we study a building block for the configuration that leads to the scattering amplitude. 

Let us consider the cusp as described in \cite{Alday:2007hr}. 
For $AdS_5$ we have that in (\ref{eq:Sg})
$q^2={r^2}/{R^2}$ and $p^2={R^2}/{r^2}$.
%\begin{equation}
%q^2=\frac{r^2}{R^2}, \qquad p^2=\frac{R^2}{r^2}.
%\end{equation}
Moreover, if we consider an ansatz of the type $ r=e^\tau w(\tau)$, 
then we find that the action is:
\begin{eqnarray}
S&=& \frac{R^2}{2\pi \alpha'}\int d\sigma d\tau \frac{1}{w(\tau)^2}\sqrt{1-(w(\tau)+\dot{w}(\tau))^2}\, .
\label{eq:AMS}
\end{eqnarray}

\begin{figure}
  \includegraphics[height=.24\textheight]{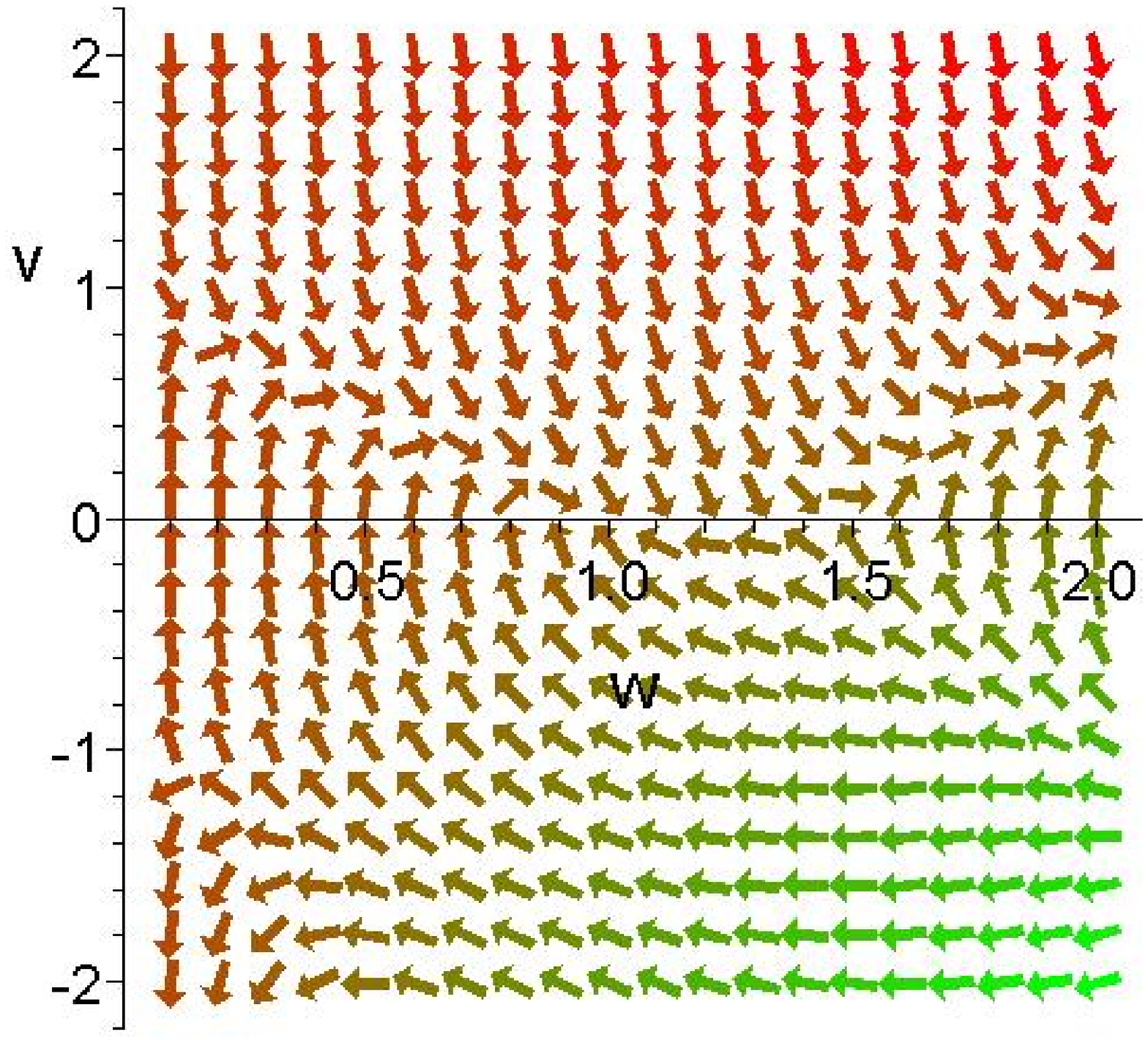}
  \includegraphics[height=.24\textheight]{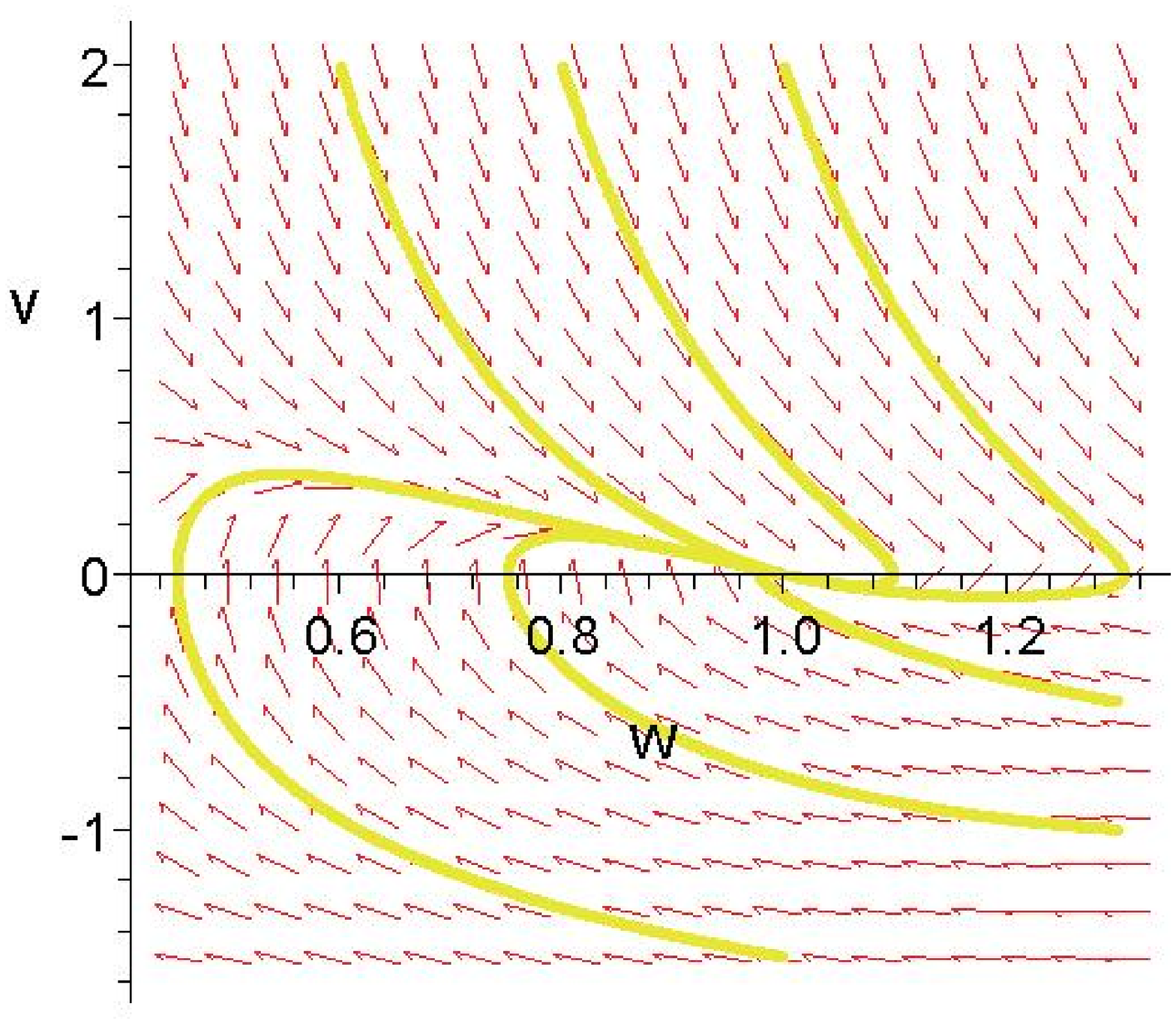}
  \includegraphics[height=.24\textheight]{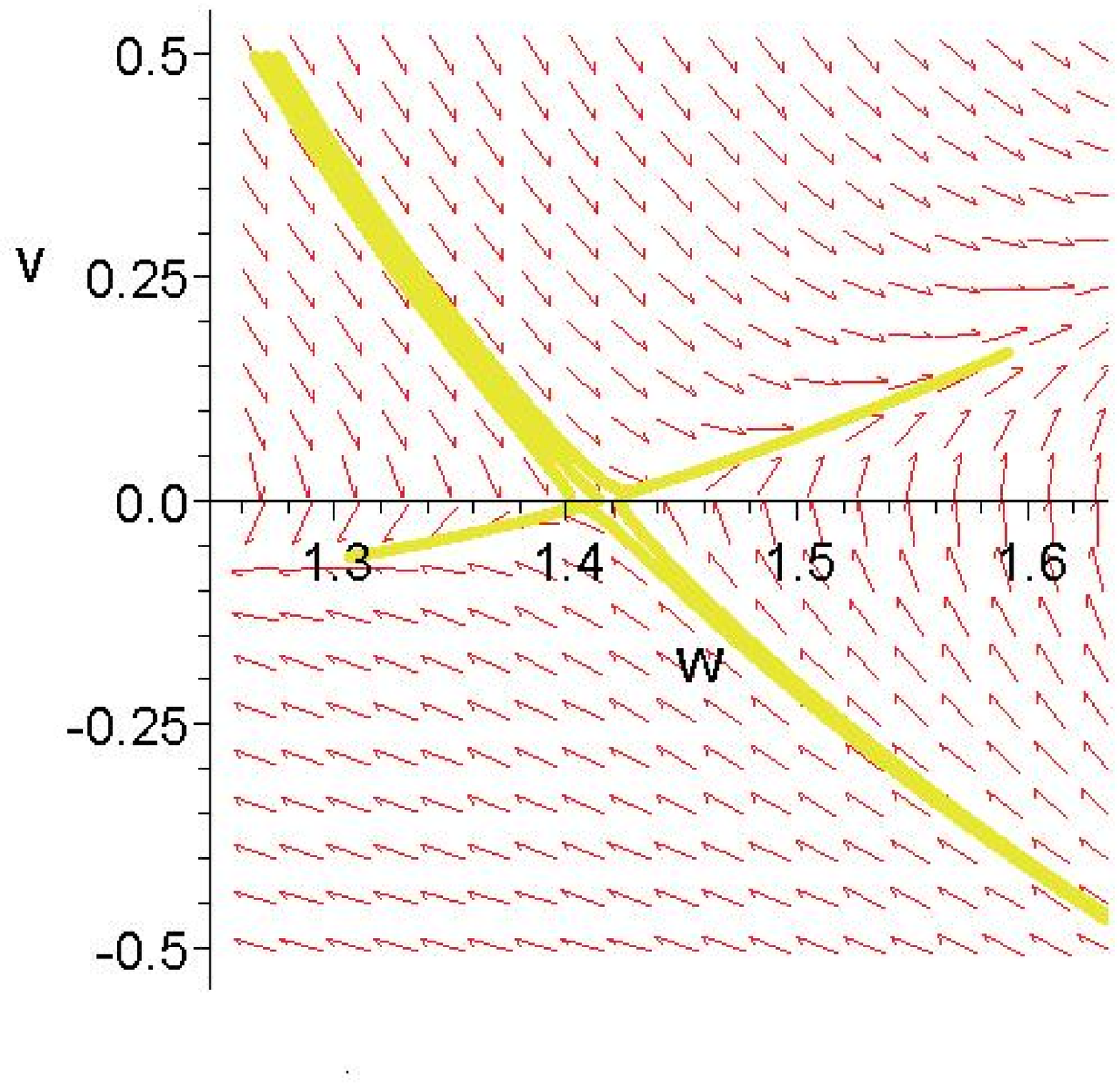}
  \caption{Left: The flow defined by system (\ref{eq:AMsys}).
           Center: The flow near the stable node.
           Right: The flow near the saddle point.}
\label{fig:AMpp}
\end{figure}

Following the methodology listed at the end of the previous subsection we derive
the corresponding equation of motion which, for the purpose of numerical integration, is better to write as the 
following system of first order ordinary differential equations,
\begin{eqnarray}
\dot{w}=v, \, \qquad 
\dot{v}=\frac 1 w \left(2-4vw-2v^2-3w^2+w^4+vw^3-v^2w^2-v^3w\right) \, .
\label{eq:AMsys}
\end{eqnarray}
Here dot stands for derivative with respect to $\tau$.
%\begin{equation}
%w w''+ww'^3+w^2 w'^2 +2 w'^2 +4w'w - w'w^3+2w^2 +w^4-2=0.
%\end{equation}
We solved this system using the seventh-eight order
continuous Runge-Kutta method. Thanks to its adaptive scheme, this method provides
a great control upon the output accuracy. 
Typical solutions are plotted in
fig.\ref{fig:AMpp}. 

An important point here is that numerical solutions are not necessarily black boxes. 
Qualitative and asymptotic analysis help to understand the numbers which in turn might guide the analysis.
For instance, in this case we can use the theory of dynamical systems to understand the contents of figures \ref{fig:AMpp}.
This theory includes a set of general techniques for the analysis of phase spaces that allows 
us to understand better the asymptotics of solutions and their dependence on the initial conditions. 
This is a natural way of tackling problems about classical solutions in string theory; 
we are not aware of previous works emphasizing these systematic methods; most of the literature emphasizes 
exploiting symmetries. While of paramount importance in $AdS_5\times S^5$, most symmetries are lacking in the context of 
supergravity backgrounds dual to nonconformal theories.  

Setting the vector field to zero in system (\ref{eq:AMsys}), we
see that there are two fixed points. 
They are respectively located at $(1,0)$ and $(\sqrt{2},0)$. 
If these singular points were hyperbolic, then the Hartman-Grobman theorem 
\cite{Guckenheimer}
ensures that, in a small neighborhood of each hyperbolic fixed point, the flow defined by the full nonlinear system (\ref{eq:AMsys}) is topologically equivalent to its linearized version.
A singular point is hyperbolic if the eigenvalues of the Jacobian of the linearization around the point have non-zero real parts.
For the first point we obtained $\lambda_1=-1$ and $\lambda_1=-2$. This indicates that in its neighbourhood the flow behaves like near a radial sink (stable node), as it is shown in the center of fig.\ref{fig:AMpp}.
For the second singular point $\lambda_1=-1+\sqrt{3}$ and $\lambda_1=-1-\sqrt{3}$.
So, in its neighborhood the flow behaves like near a saddle point, as shown to the right in fig.\ref{fig:AMpp}.

Having tested that the numerical and analytical results coincide, we can move forward and integrate (\ref{eq:AMS}) numerically.
Taking $\Lambda=r_{max}$, we evaluate the action with a set of different upper limits $t_{max}$ which are defined by $r_{max}$.
Then using $\frac{d}{d\Lambda} = \frac 1{\dot{r}_{max}} \frac{d}{dt_{max}}$,  
we estimate the cusp anomalous dimension using (\ref{eq:cusp}).
The outcome for this case nicely replicates the result in \cite{Alday:2007hr}. 
Because of lack of space, we refrain of presenting it here, and move forward to the nonconformal cases.

%%%%%%%%%%%%%%%%%%%%%%%%%%%%%%%%%%%%%%%%%%%%%%%%%%%%%%%%%%%%%%%%%%  
\section{Cusp anomalous dimensions in nonconformal gauge theories}
\label{sec:cuspNC}

%%%%%%%%%%%%%%%%%%%%%%%%%%%%%%%%%%%%%%%%%%%%%%%%%%%%%%%%%%%%%%%%%%  
\subsection{Witten QCD} \label{ssec:WQCD} 
%%%%%%%%%%%%%%%%%%%%%%%%%%%%%%%%%%%%%%%%%%%%%%%%%%%%%%%%%%%%%%%%%%%  

In this subsection we calculate the cusp anomalous dimension for the Witten background \cite{Witten:1998zw}. 
With that aim we rewrite the metric in the relevant IR regime.
The ten-dimensional string frame metric and
dilaton of this model are 
\begin{eqnarray}
\label{defns} 
ds^2&=&({u\over
R})^{3/2} (\eta_{\mu\nu}dx^\mu dx^\nu + {4R^3\over
9u_0}f(u)d\theta^2)+ ({R\over u})^{3/2}{du^2\over f(u)}
+R^{3/2}u^{1/2}d\Omega_4^2\ ,\nonumber \\    
f(u)&=&1-{u_0^3\over
u^3}\ , \qquad \qquad R=(\pi
Ng_s)^{1\over3}{\alpha'}^{1\over2}\, , \qquad      
e^\Phi=g_s{u^{3/4}\over R^{3/4}}\ . 
\end{eqnarray}

The geometry consists of a warped, flat 4-d part, a radial direction $u$, a circle parameterized by $\theta$ with radius vanishing at the horizon $u=u_0$, and a four-sphere whose volume is instead everywhere non-zero.
It is non-singular at $u=u_0$.
In the $u\to\infty$
limit the dilaton diverges: this implies  that in this limit the
completion of the present IIA model has to be found in M-theory.
The background is completed by a constant four-form field strength  
%\begin{equation}
%F_4=3R^3\omega_4\ ,  
%\end{equation}
$F_4=3R^3\omega_4$,
where $\omega_4$ is the volume form of the
transverse $S^4$.  

The main gauge theory parameter we will use in the following is the KK mass scale
$1/R_{\theta}=3m_0/2$, where $m_0^2={u_0}/{R^3}$.    
%\begin{eqnarray}
%\label{emmezero}
%\frac{1}{R_\theta}=\frac32m_0\ ,\qquad {\rm where}\qquad
%m_0^2=\frac{u_0}{R^3}\ .    
%\end{eqnarray}    
As can be read from the metric,
$m_0$ is also the typical glueball mass scale, and its square is proportional to the ratio between the
confining string tension $T_{QCD}$ and the UV 't Hooft coupling
$\lambda$.  
The supergravity approximation works in the regime
opposite  to that in which the KK degrees of freedom decouple from
the low energy dynamics.
Condition $T_{QCD}\ll m_0^2$ implies $\lambda\ll 1$, which
is beyond the supergravity regime of validity.

\begin{figure}
  \includegraphics[height=.23\textheight]{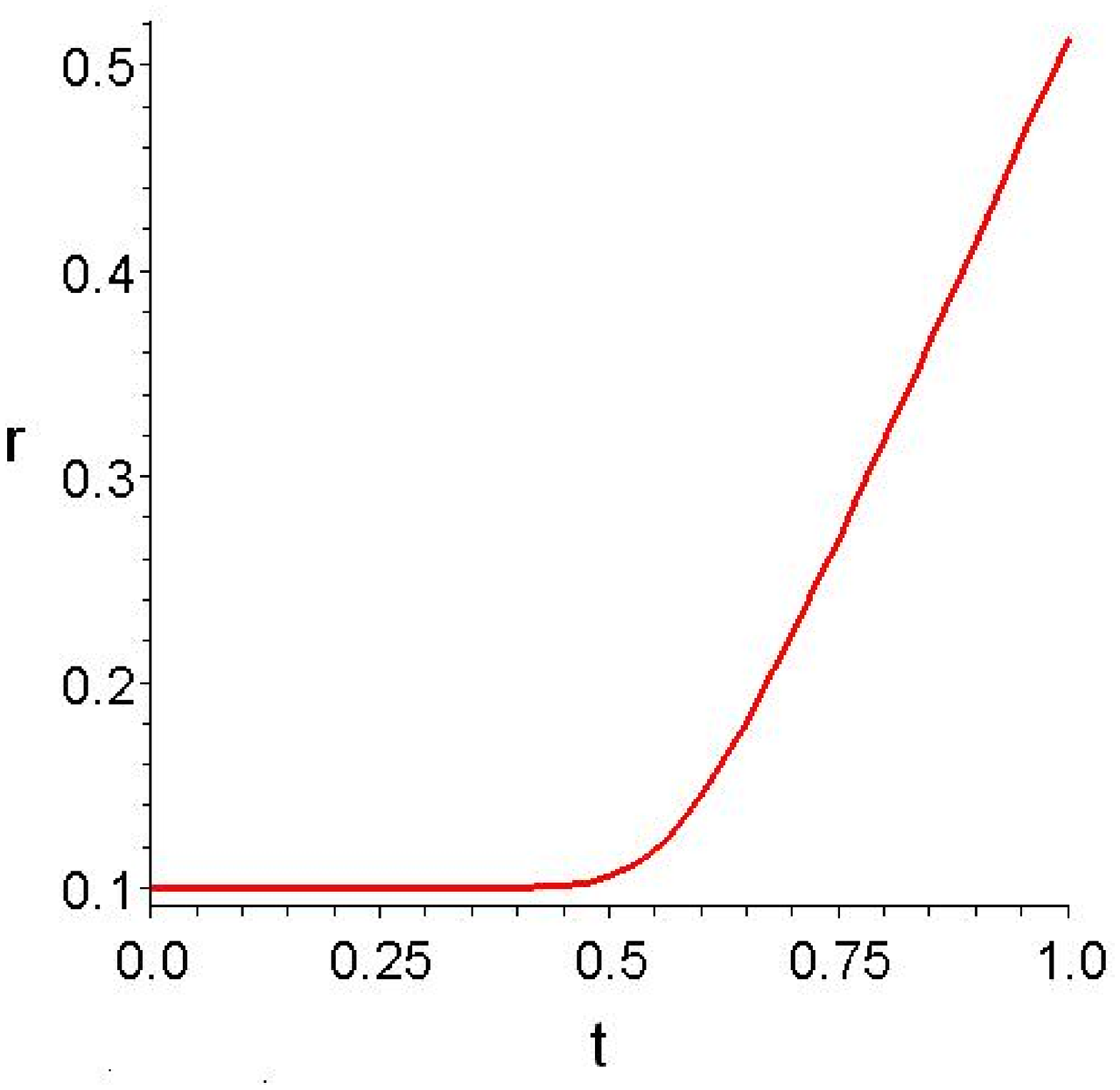}
\includegraphics[height=.23\textheight]{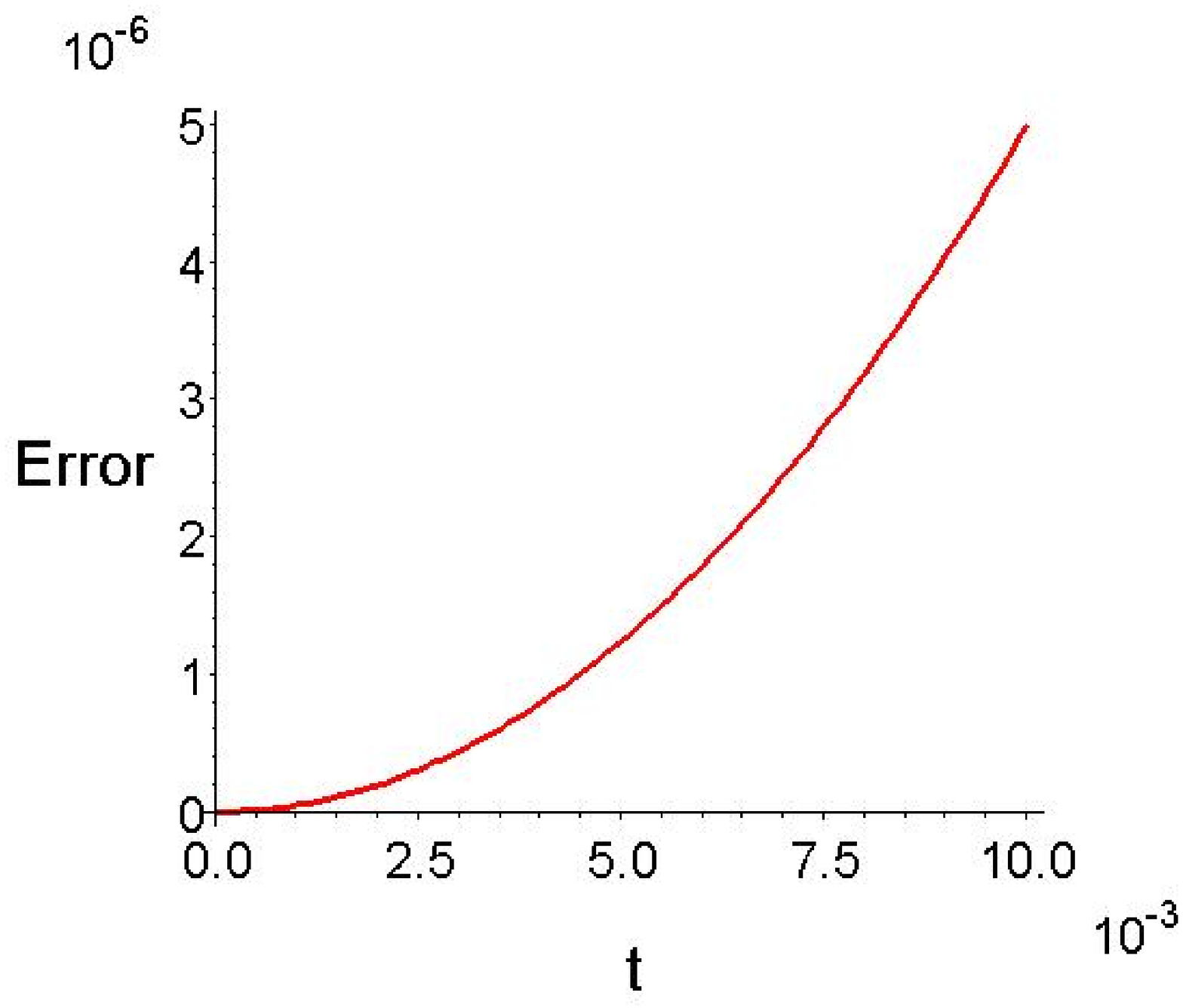}
\includegraphics[height=.23\textheight]{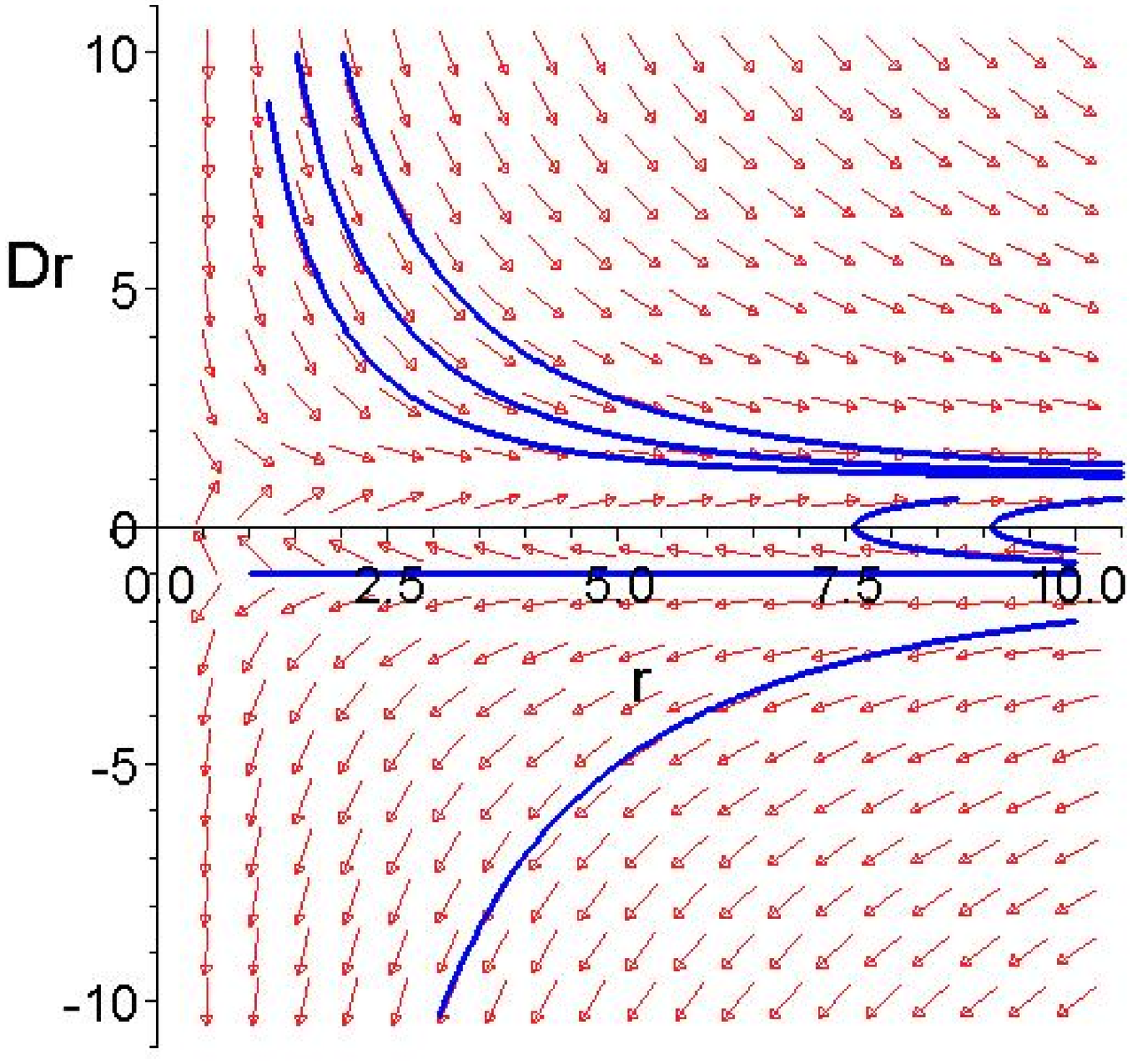}
  \caption{Left: a typical solution for $r$ in the WQCD model. Center: error of the approximate solution (\ref{eq:smalltr}) compared to the numerical solution. Right: phase space for very large $t$. $Dr$ stands for the derivative of $r$ with respect to $t$.}
  \label{fig:WQCDsolr}
\end{figure}

Assuming that $u(\tau,\sigma)=r(\tau)$, with
$q(r)^2=\left({u}/{R}\right)^{3/2}$, 
$p(r)^2 = \left({R}/{u}\right)^{3/2} \times 1/{f}$,
and introducing the new coordinate $t=e^\tau$, so that $e^{-\tau}\partial/\partial \tau =\partial/\partial t$, the 
action (\ref{eq:Sg}) is now:  
\begin{equation}
S=\frac{R^{3/2}}{2\pi \alpha'}\int d\sigma\, \int dt \,\,t \frac{1}{r^{3/2}} 
\sqrt{1-(\partial_t r)^2/f} .
\end{equation}
A convenient way to write the equation of motion is:
\begin{equation}
\ddot{r} = \frac{1}{2} \frac{\left[2r^7\dot{r}^3-3tr^3(r^3-2r_0^3)\dot{r}^2-2r^4(r^3-r_0^3)^2\dot{r}+3t(r^3-r_0^3)^2\right]}{r^4 t(r^3-r_0^3)} \, ,
\label{eq:WQCDeom}
\end{equation}
where $r$ and $t$ are now measured in units of $R$, and a dot stands for derivative with respect to $t$.
Note that $r=const=r_0$ is not properly a fixed point of the action. 
A typical solution of this second order non-linear non-autonomous equation is presented in the left of fig.\ref{fig:WQCDsolr}.
As with the rest of the examples plotted in this subsection, 
it was obtained by setting $r_0=1$, $t_0=10^{-6}$, $r(0)=1+10^{-6}$ and $\dot{r}(0)=0$. 
To understand this result let us analyze the asymptotic regimes.

\subsubsection{Asymptotic behavior of the solution}

First, we look for a solution for $(t-t_0) \rightarrow 0$ and $\epsilon = r(0)-r_0 \rightarrow 0$.  
Balancing equation (\ref{eq:WQCDeom}) the relevant terms are,
\begin{equation}
 \ddot{r} =
\frac{r^3 \dot{r}^3-(r^3 - r_0^3)\dot{r}}{t(r^3-r_0^3)} \, .
\label{eq:smallt}
\end{equation}
For this balance to be
consistent it is required the absolute value of the second
derivative to be greater than the neglected terms. 
It can be readily tested that this condition is fulfilled in the limits of $\epsilon$ and $t$ we are considering.

We inmediately note that any solution to (\ref{eq:smallt}) with initial condition $\dot{r} = 0$ at $t_0 \neq 0$ is a constant solution. 
A solution to this equation given as an expansion in Taylor series is,
\begin{equation}
r =
(r_0+\epsilon)+\epsilon(t-t_0)
+\frac{\epsilon(r_0^3\epsilon+3r_0^2\epsilon^2+3r_0\epsilon^3+\epsilon^4-3r_0^2-3r_0\epsilon-\epsilon^2)}{2t_0(3r_0^2+3r_0\epsilon+\epsilon^2)} 
\dots
\label{eq:smalltr}
\end{equation}
It can be seen that, 
in the limit $\epsilon \rightarrow 0$,
$r_0$ actually acts as a fixed point of the system.
We tested this result numerically.
In the center of fig.\ref{fig:WQCDsolr}
%\begin{figure}
%  \includegraphics[height=.3\textheight]{smalltError}
%  \caption{Plot of the error of the approximate solution (\ref{eq:smalltr}) compared to the numerical solution with $t_0=10^{-6}$.}
%\label{fig:smalltError}
%\end{figure}
we can see that the difference between (\ref{eq:smalltr}) and a corresponding numerical solution is negligible.
Thus, the approximation works in the range of $t$ and $\epsilon$ considered.

Let us now consider the asymptotic behavior of the solution for $t \rightarrow \infty$
Now the equation of motion (\ref{eq:WQCDeom}) becomes
\begin{equation}
\ddot{r}+ \frac{3}{2} \frac{\dot{r}^2
(r^3-2 r_0^3)}{(r^3-r_0^3)r}
-\frac{3}{2}\frac{r^3-r_0^3}{r^4}=0 \, .
\label{eq:larget}
\end{equation}
With regards to the consistency of this balance, from fig.\ref{fig:WQCDsolr} it can be assumed that the large $t$ behavior of the solutions is linear. 
Then, the second derivative goes to zero 
and that is also true for the
neglected terms, $
(r^3\dot{r}^3)/(t(r^3 - r_0^3)) - \dot{r}/t\, .$
For further verification we again compare the analytical and numerical results.
Note now that the second derivative is zero when the first derivative is
$\pm (r^3-r_0^3)((r^4-2r_0^3 r)^{1/2}r)\, $ whose limit as 
$r\rightarrow\infty$ gives $\pm 1$.
So, for any $r_0$, some solutions converge to a linear solution with
slope $1$ and some diverge from a linear solution with
slope $-1$. 
In the right of fig.\ref{fig:WQCDsolr}
%\begin{figure}
%  \includegraphics[height=.3\textheight]{largetpp}
%  \caption{The phase space describing the behaviour of $r$ for very large $t$. $Dr$ stands for the derivative of r with respect to time.}
%\label{fig:larget}
%\end{figure}
the phase-space of (\ref{eq:larget}) is plotted.
We see that, indeed, for any $r_0$,  $\dot{r}=1$ is an
attractive set and that the solutions diverge from $\dot{r}=-1$.
Now, recall that we are analyzing the behavior for $t\rightarrow
\infty$ and, according to the analysis of the solutions when $t
\rightarrow 0$, the interesting us region is $r>r_0$ and
$\dot{r}>0$. It means that the interesting us asymptotic
behavior is given by the linear solution with slope equal to $1$.

Summarizing, the analysis and the numerical results for the range of $r_0$ tested ($r_0 \in [10^{-6},10]$ and $ t_0=10^{-6}$)
indicate that the generic behavior of the solutions is as follows:
$r_0$ acts as a fixed point of this system; 
with an initial condition near $r_0$, the solution starts with a regime where $r \approx r_0$.
Next there is a transient regime where the solution smoothly changes
until it finally sets into a linear behavior with unit slope. The
closer the initial condition to the horizon, the larger the length
of the initial regime and the shorter the transition regime. The
length of the time intervals are to be considered relatively, since
the figures show always the same behavior if the right lengths of
the $t$ interval are chosen.

\subsubsection{The cusp anomalous dimension}
\begin{figure}
  \includegraphics[height=.23\textheight]{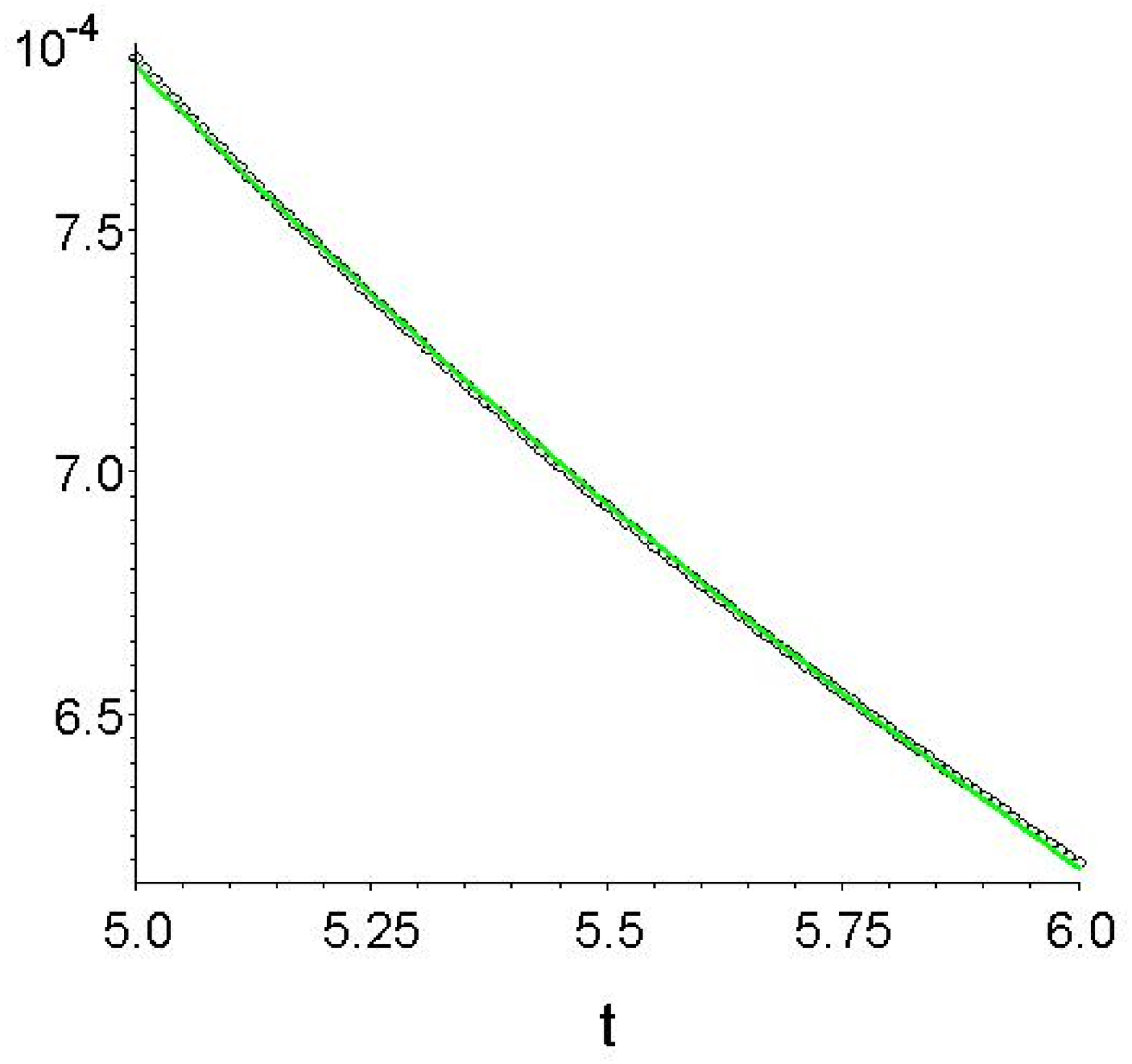}
 \includegraphics[height=.23\textheight]{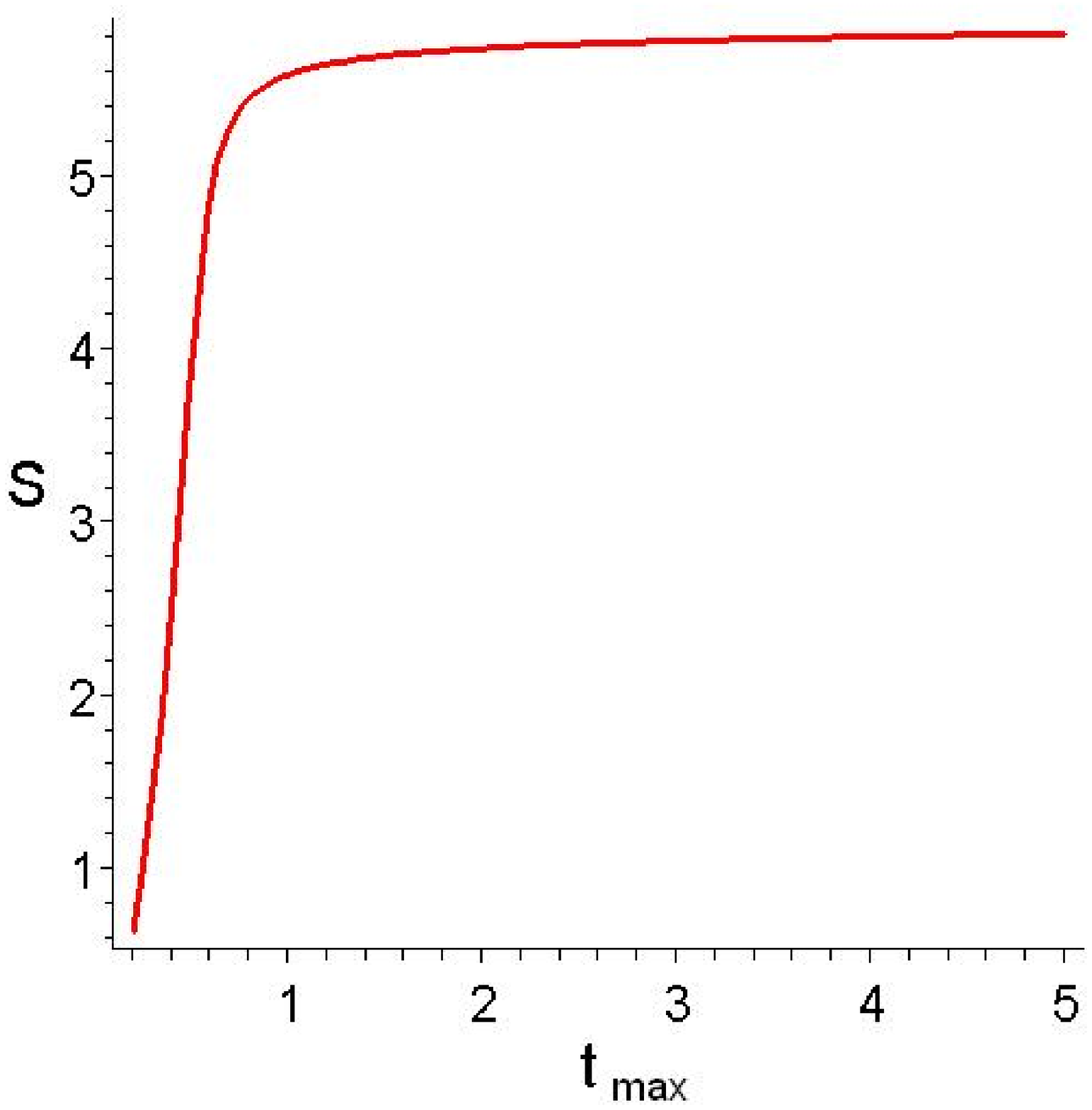}
\includegraphics[height=.23\textheight]{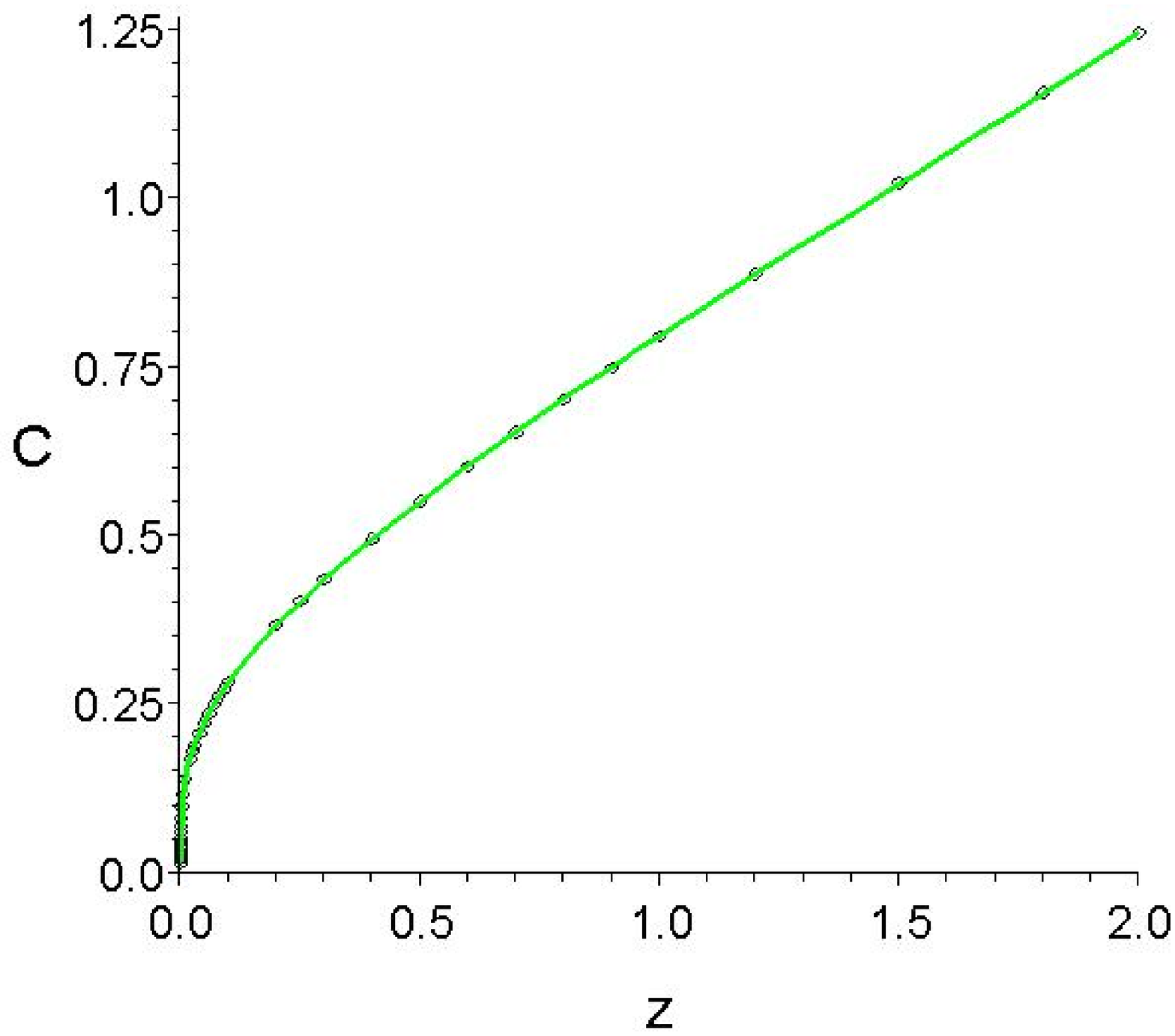}
  \caption{Left: fit to $1/t$ of the argument of the squared root in the WQCD action. 
Center: plot of the integral as function of $t_{max}$.
Right: fit of C as function of z.}
\label{fig:WQCDaction}
\end{figure}

Next, following our methodology, 
we evaluate the numerical solution in the action and integrate it numerically.
To accurately compute the actions we used an adaptive 3-5 Simpson's quadrature.
If numerical factors are not taken into account, the integrand is,
$
\frac{t}{r^{3/2}}\sqrt{\frac{r^3(1-\dot{r}^2)-1}{r^3-1}} \, .
$
We have already seen that $r\rightarrow t$ while $t \rightarrow \infty$.
Therefore, if we expect this integral to diverge logarithmically,
we need the argument of the squared root to behave like $1/t$.
Indeed, using a best fit to the numerical results, 
we have verified that this is the case, and we present an example in the left side of fig.\ref{fig:WQCDaction}.
In the center
%Finally, in fig.\ref{fig: Svst}
%\begin{figure}
%  \includegraphics[height=.3\textheight]{Svst}
%  \caption{Plot of the action as function of time.}
%\label{fig: Svst}
%\end{figure}
a typical result for the action as a function of $t_{max}\propto r_{max} = \Lambda$ is plotted. 

The coefficient of $dS/dlnr_{max}$ is $A \sqrt{C}$,
where $A$ stands for the numerical factors in front of the integral, $C=
r_0 (b z^p + b_0 + b_1 z + b_2 z^2 + b_3 z^3)$, $z = r(0)/r_0 -1$
and $p=0.27895, b = 0.436898, b_0 = 0.016228, b_1 = 0.359128, b_2 =
-0.026669, b_3 = 0.01103$. 
The numbers were obtained by the best fit $C(z)$
shown in the right of fig.\ref{fig:WQCDaction}.
%\begin{figure}
%  \includegraphics[height=.3\textheight]{Cvsz}
%  \caption{Fit of C as function of z.}
%\label{fig:Cvsz}
%\end{figure}

%From this equation it follows that we can have $\dot{r}$ and $\ddot{r}$ vanishing if we have $r=r_0$. Thus, $r_0$ is a 
%fixed point of the equations of motion. The precise analysis of what kind of fixed point is still to be carried out. 

%\noindent
%$\bullet$ This solution might not satisfy the boundary conditions we require.

%\noindent
%$\bullet$ Possible solutions to consider: 
%\[
%r=r_* + \sum a_n t^n.
%\]
%This solution has the advantage that is satisfies the boundary conditions naturally. Namely, for $r=r_*$ we have that 
%$t=e^\tau=\sqrt{y_0^2-y_1^2}=0$ and thus we are in the light cone. The main disadvantage of this solution is convergence 
%for large $t$. 

%The other solution that we can consider is:
%\[
%r=r_* + r_0^n \beta_n(t).
%\]
%The zeroth order solution is known and is presented in \cite{Alday:2007hr}. 
%It has the advantage of allowing us to compute is regime of validity by considering $t$ such that the corrections are of 
%the same order as the zeroth order solution solution $r_*$. The disadvantage is that $r_0$ is related to the glueball mass 
%in the field theory and very small glueball mass is not a necessarily clean limit. 

\subsection{The warped deformed conifold}

The Klebanov-Strassler background \cite{Klebanov:2000hb} is very well known in the literature of gravity duals 
to confining gauge theories, a full presentation of its metrics and action, suited for numerical 
calculations can be found in the appendix A of \cite{Mahato:2007zm}.

\begin{figure}
\includegraphics[height=.3\textheight]{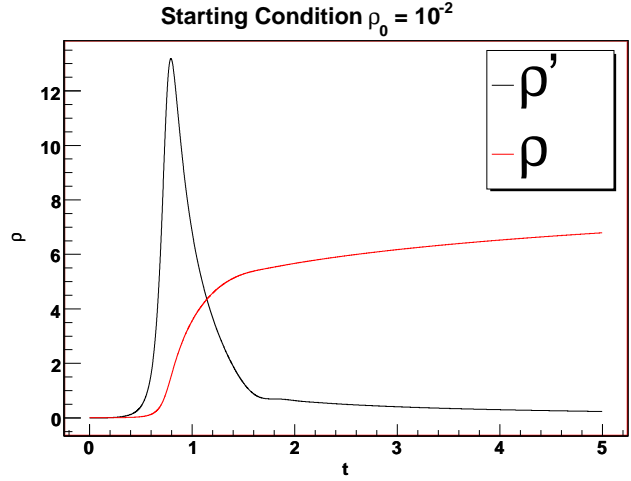}
\caption{There is a fixed point of the Klebanov-Strassler equations of motion at $\rho_0 = 0$.  This plot is for $\epsilon = 10^{-2}$. }
\label{fig:r0-2}
\end{figure}

Using coordinates where $t=e^{\tau}$ and setting $\rho(t,\sigma)=\rho(t)$,
the equation of motion for this background is 
\begin{eqnarray}
&&\frac{d}{dt} \left( -\frac{\epsilon^{4/3}h^{1/2}(\rho)\dot{\rho}}
{6t K^2(\rho)\sqrt{1-\frac{\epsilon^{4/3}}{6t^2 K^2(\rho)}\dot{\rho}^2}}  \right)= \\
&&t h^{1/2}(\rho) \sqrt{1-\frac{\epsilon^{4/3}}{6t^2 K^2(\rho)}\dot{\rho}^2} \left( \frac{\partial_{\rho} h(\rho)}{2h(\rho)} - \frac{\epsilon^{4/3} \dot{\rho}^2 \partial_{\rho}K(\rho)}{t^2 K^3(\rho)(1-\frac{\epsilon^{4/3}}{6t^2 K^2(\rho)} \dot{\rho}^2)} \right) \nonumber \, .
\end{eqnarray}
This equation was solved by setting $\rho(0) = \epsilon \rightarrow 0$ and $\rho^{\prime}(0) = 0$, and then 
using a fourth order Runge-Kutta method for the evolution of the solution as a function of $t = e^{\tau}$. The 
numerical integration necessary for evaluating the warp factor $h(\rho)$ was done using a composite Simpson method.
A typical solution is presented in fig.\ref{fig:r0-2}.
An analysis similar to the previous subsections shows that there is a fixed point at $\rho = 0$. 
The solution initially stays close to 
the fixed point, then there is a transition to a logarithmic looking function.  The region 
close to the fixed point is larger depending on how close to the fixed point the 
solution begins.  

To show that the 
%large $t$ 
solution grows logarithmically, we examine the form of the metric for large $t$.  
In this regime, $\rho$ is also getting large, so we have 
$K(\rho) \to B\,\exp(-\rho/3)$,
%\begin{equation}
%K(\rho) \to B e^{-\rho/3} \, ,
%\end{equation}
where $B$ is a constant.  The KS metric now has the form
$
ds^2 = h^{-1/2}(\rho) dx^2 + h^{1/2} e^{2\rho/3} d\rho^2.
$
Substituting $r = e^{\rho/3}$ and performing a T-duality on the $x$ coordinates, the metric is 
\begin{equation}
ds^2 = h^{1/2}(3 \log{r}) (dx^2 + dr^2).
\end{equation}
Now, in the large $r$ (large $\rho$) approximation $h(\rho) \to A (3 \log{\rho} - 1/4)/r^4,$ 
where $A$ is a constant, and one gets 
\begin{equation}
ds^2 =  \sqrt{3 \log{\rho} - 1/4}\frac{dx^2 + dr^2}{r^2}
\end{equation}
which is $AdS_5$ with a logarithmic prefactor.  To a first approximation, 
we can ignore the prefactor which will change slowly, to the metric, to see that 
the cusp solution of $AdS_5$ which is $r = constant \times t$ will hold, which means that $\rho \sim \log{t}$ for large $t$.

This way, the result for KS resembles very much the one obtained for WQCD.

\section{Discussion and outlook}
In this contribution we have shown that light-like Wilson loops in nonconformal theories 
display some rather universal features. We hope that this is the kind of question that can be asked and 
whose answer should be trusted to be in the same universality class as QCD. The light-like Wilson loops we have considered 
here are building blocks of the scattering amplitudes. We hope, as is the case in the 
perturbative treatment of QCD amplitudes, that the main contribution to scattering amplitudes will come from the 
cusp region and therefore, our calculations will be a crucial building block in the amplitudes. 

It is important to note that, although not included in this contribution, we have performed similar analysis for various 
theories and the behavior is rather similar to the one observed in the two examples considered explicitly here. Basically, there 
is a natural infrared regulator which is characterized by the distance between the position of the brane and the characteristic 
strong scale which in most supergravity solutions is given by the region where a
cycle collapses ($r_0$ for WQCD and $\rho=0$ for KS). 
We want to emphasize that in the perturbative approach to gauge theories the infrared regulator normally used is formal and has 
nothing to do directly with the strong scale. We thus hope that our strong coupling calculation will help shed some light 
into the structure of IR singularities.

%%%%%%%%%%%%%%%%%%%%%%%%%%%%%%%%%%%%%%%%%%%%%%%%
%% BACKMATTER
%%%%%%%%%%%%%%%%%%%%%%%%%%%%%%%%%%%%%%%%%%%%%%%%

\begin{theacknowledgments}
First, we wish to thank the organizers of the workshop "Ten years of AdS/CFT" for the invitation to present our results in this important meeting. 
CAT-E is grateful to the CEFIMAS and the MCTP for hospitality while part of this research was undertaken. LPZ and DJP are partially supported by 
DoE under grant DE-FG02-95ER40899.
\end{theacknowledgments}

%%%%%%%%%%%%%%%%%%%%%%%%%%%%%%%%%%%%%%%%%%%%%%%%
%% The bibliography can be prepared using the BibTeX program or
%% manually.
%%
%% The code below assumes that BibTeX is used.  If the bibliography is
%% produced without BibTeX comment out the following lines and see the
%% aipguide.pdf for further information.
%%
%% For your convenience a manually coded example is appended
%% after the \end{document}
%%%%%%%%%%%%%%%%%%%%%%%%%%%%%%%%%%%%%%%%%%%%%%%%

%%%%%%%%%%%%%%%%%%%%%%%%%%%%%%%%%%%%%%%%%%%%%%%%
%% You may have to change the BibTeX style below, depending on your
%% setup or preferences.
%%
%%
%% For The AIP proceedings layouts use either
%%%%%%%%%%%%%%%%%%%%%%%%%%%%%%%%%%%%%%%%%%%%

\bibliographystyle{aipproc}   % if natbib is available
%\bibliographystyle{aipprocl} % if natbib is missing

%%%%%%%%%%%%%%%%%%%%%%%%%%%%%%%%%%%%%%%%%%%
%% You probably want to use your own bibtex database here
%%%%%%%%%%%%%%%%%%%%%%%%%%%%%%%%%%%%%%%%%%%
\bibliography{sample}

%%%%%%%%%%%%%%%%%%%%%%%%%%%%%%%%%%%%%%%%%%%
%% Just a reminder that you may have to run bibtex
%% All of it up to \end{document} can be removed
%% if you don't like the warning.
%%%%%%%%%%%%%%%%%%%%%%%%%%%%%%%%%%%%%%%%%%%
\IfFileExists{\jobname.bbl}{}
 {\typeout{}
  \typeout{******************************************}
  \typeout{** Please run "bibtex \jobname" to optain}
  \typeout{** the bibliography and then re-run LaTeX}
  \typeout{** twice to fix the references!}
  \typeout{******************************************}
  \typeout{}
 }

\end{document}